# Role of Exchange in Density Functional Theory for Weakly-Interacting Systems: Quantum Monte Carlo Analysis of Electron Density and Interaction Energy

Yosuke Kanai and Jeffrey C. Grossman

Berkeley Nanosciences and Nanoengineering Institute
University of California, Berkeley.

**ABSTRACT**

We analyze the density functional theory (DFT) description of weak interactions by employing diffusion and reptation quantum Monte Carlo (QMC) calculations, for a set of benzene-molecule complexes. While the binding energies depend significantly on the exchange correlation approximation employed for DFT calculations, QMC calculations show that the electron density is accurately described within DFT, including the quantitative features in the reduced density gradient. We elucidate how the enhancement of the exchange energy density at a large reduced density gradient plays a critical role in obtaining accurate DFT description of weakly-interacting systems.

Weak interactions play an important role in numerous chemical, physical and biological phenomena in nature [1], and vast opportunities exist for using weak interactions for various technological applications such as hydrogen storage for renewable energy and highly selective coatings for bio-chemical detectors [2,3]. Our ability to accurately describe such interactions in theoretical calculations is important for advancing these technologically important fields.

Density functional theory (DFT) [4,5] is a promising method for describing the electronic structure of realistic systems because of its applicability to a large class of materials ranging from molecules to solids, in terms of both accuracy and computational affordability. Weakly-interacting systems, however, remain a challenging class of materials to describe accurately within the DFT approaches in practice [6]. The difficulty has been attributed primarily to the dominant role of nonlocal correlation in describing weak interactions such as the van der Waals interaction, which is absent or incorrectly accounted for within many exchange-correlation (XC) approximations. There have been a number of efforts to either empirically or formally include nonlocal *correlation* in the XC approximation [7]. In addition to this, a quantitative description remains highly challenging due to the pairing *exchange* part, which requires further investigation [8] and is in general considerably larger than the correlation part.

In the context of improving the accuracy of DFT, quantum Monte Carlo (QMC) calculations have played an important role in the development of the XC approximation, starting with the seminal work of Ceperley and Alder on the homogeneous electron gas [9]. With computational and methodological advances, it is now becoming possible for QMC to compute accurate electron densities for realistic systems. In this Rapid Communication, we employ QMC calculations to analyze the electron density and binding energies calculated from DFT in order to elucidate the role of *exchange* in the XC approximation for describing weak interactions. In spite of the severe XC approximation dependence of the binding energy, our QMC results show that both the electron density and the reduced density gradient (RDG) are described quite accurately by DFT. Using these results, we show that an enhancement of the exchange energy density at large RDG values play a critical role in obtaining accurate binding energies. We demonstrate that the diverging behavior of this enhancement factor at large RDG among different exchange approximations leads to significant differences in the binding energy. Taken together, these results show that the exchange description in XC approximations needs to be improved if DFT is to describe quantitatively and correctly the physics of weakly-interacting systems, even with an accurate inclusion of nonlocal correlation. Tailoring the exchange enhancement factor at large RDG for weak interactions might improve significantly the description while essentially leaving unaffected other types of interactions and avoiding the computationally expensive optimized effective potential approach to obtain the exact exchange.

DFT calculations were performed as implemented in the GAMESS code [10]. Pseudopotentials [11] were used to describe core electrons for all atoms and optimized Gaussian basis sets of triple-zeta+polarization quality were used. Basis set superposition error to the binding energy was estimated to be less than 0.005 eV using the counterpoise method [12] for the benzene-$H_2$ complex, negligibly small for the present discussion and thus not included in reported values. All equilibrium geometries are determined such that forces are smaller than $1.0 \times 10^{-5}$ eV/Å, and the binding energies are computed using the structures relaxed within the given XC functional.

Fixed-node diffusion QMC calculations are employed in order to obtain highly accurate binding energies [13]. The calculation is based on solving the imaginary-time Schrödinger equation, $-\partial_\tau \Psi(R,\tau) = (\hat{H} - E_0)\Psi(R,\tau)$ which yields the many-body eigenstate $\Phi_0$ with the eigenvalue $E_0$ as the imaginary-time $\tau$ goes to infinity. In practice, the integral form of this equation is employed with the short-time approximation of the Green's function with a time step of 0.01 a.u. Importance sampling is introduced with trial wavefunctions $\Psi_T$, which are obtained using a variational QMC calculation from a Slater determinant of Kohn-Sham orbitals, multiplied by a two-body Jastrow correlation factor. Pseudopotentials are used to describe the core electrons [11], and the QWalk code was used for all QMC calculations [14].

The binding energy computed using several XC approximations within DFT [15] and diffusion QMC is shown for a set of weakly-interacting benzene-molecule complexes (Figure 1). While most functionals yield similar separation distances, varying only marginally by ~ 0.1 Å when bound, LDA generally results in a much shorter separation distance, differing considerably from others by as much as ~ 0.4 Å for some cases. It is evident that the DFT binding energy for all complexes varies significantly depending on which XC approximation is employed. LDA significantly overbinds all the complexes in general, compared to the binding energies determined by QMC calculations at LDA geometries (QMC-LDA). However, the observed *trend* for the set is consistent with the QMC-LDA calculations. PBE values appear to be rather close and consistently smaller than QMC-PBE (QMC at PBE geometries) values in all cases, and both PW91 and PBE0 closely follow the same trend (these two functionals give essentially the same geometries as PBE). Another interesting observation is that those functionals with the B88 exchange (BP86, BLYP and B3LYP) are all significantly *under*-bound for all complexes, compared to both QMC-LDA and QMC-PBE as well as other DFT values, and with this exchange approximation no binding energies were found for half of the complexes that should be bound according to the QMC calculations. It is clear from these calculations that the XC dependence of the binding energy in DFT is unacceptably large for making meaningful predictions in many important technological applications. For example, the binding energy needed for hydrogen storage applications is on the same order of magnitude as the variation observed among the XC approximations [16].

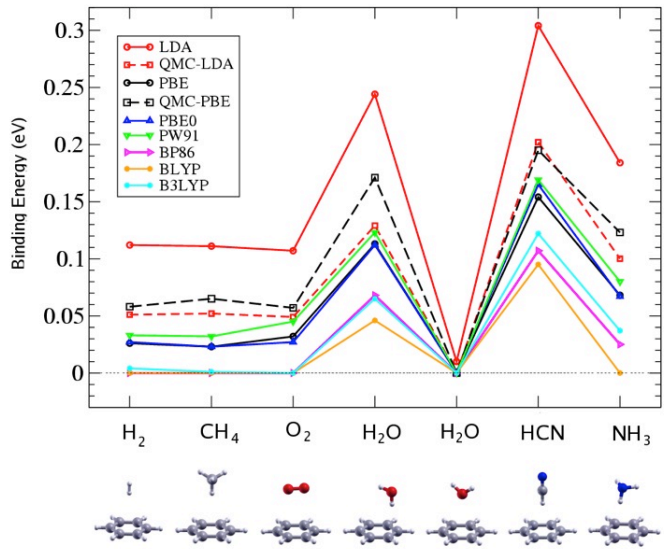

**Figure 1.** Binding energies (in eV) calculated with different XC approximations in DFT. QMC-LDA and QMC-PBE refer to the QMC binding energies calculated at the equilibrium geometries and using fixed Fermion nodes of LDA and PBE, respectively. The statistical uncertainties in QMC values are less than +/- 8 meV. Benzene-molecule complexes are shown with $H_2$, $CH_4$, $O_2$, $H_2O$ with H down (pointing toward benzene), $H_2O$ with O down, HCN, and $NH_3$ molecules.

In order to address this issue, we need first to assess the accuracy of the electron density and reduced density gradient obtained from DFT calculations for describing weak interactions. The dimensionless reduced density gradient (RDG) is defined $s(r) \equiv |\nabla n(r)|/2k_F n(r)$, where $k_F$ is the local Fermi wavevector, and $n(r)$ is the density. The electron density and its gradient are key ingredients in XC approximations for DFT calculations. Since the density operator does not commute with the Hamiltonian, the straightforward application of diffusion QMC is not possible, suffering from the mixed estimator error, and the use of the forward-walking technique can also suffer from significant statistical noise [17]. Therefore, here we calculate the electron density using the recently developed reptation quantum Monte Carlo approach [18], going beyond the variational quantum Monte Carlo in which the mathematical form of the wavefunction is fixed. Reptation QMC is based on the imaginary-time Schrödinger equation as in the case of diffusion QMC, but the random walk is performed with the Green's function $G$ in the "path" space $l = [R_0, R_1, ..., R_n]$. Sampling the path distribution,

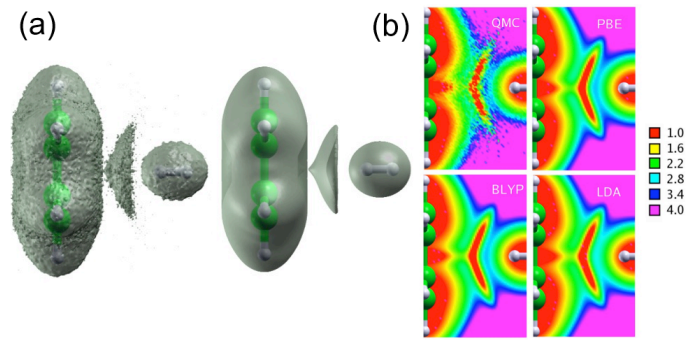

**Figure 2.** (a) Reduced Density Gradient (RDG) from R-QMC and DFT-PBE, plotted with the isosurface of 2. Note that the RDG is smaller inside the space enclosed by the isosurface. (b) The contour plot of RDG from R-QMC and DFT using PBE, BLYP, and LDA XC approximations on the $C_{6v}$ symmetry plane containing two carbon atoms.

$$\prod(l) = \Psi_T(R_0)G(R_0,R_1,\tau)...G(R_{n-1},R_n,\tau)\Psi_T(R_n),$$

the pure distribution $\Phi_0^2$ is obtained from the distribution of $R_{n/2}$, in order for the density to be calculated. A path length of 3 a.u. and a time step of 0.01 a.u. were used. Since it is computationally quite expensive to obtain the electron density accurately, we consider here only the benzene-$H_2$ complex.

Comparison of the densities at a fixed geometry [19] from QMC and DFT calculations (LDA, PBE, BLYP XC approximations) revealed a negligible difference except for the LDA density, which shows somewhat stronger overlap of densities in the middle, although with only a 2~3 % average deviation from others (never exceeding 0.3 m$e$/a.u. difference). The quantitative differences between QMC and PBE/BLYP densities are smaller than the statistical uncertainties of the QMC density in our calculation (See Supplementary Materials). A comparison of the reduced density gradient (RDG) computed from the DFT density to the one from the QMC density is shown in Figure 2. The slope of the RDG changes its sign due to the interaction, forming the valley in the middle. The RDG isosurface comparisons at a value of 2 for QMC and DFT-PBE and also to other XC approximations are shown. Within DFT calculations, PBE and BLYP are essentially identical while LDA appears to make the features slightly less pronounced. The maximum variation of LDA from PBE/BLYP in the RDG was only ~ 0.3 in this region (RDG < 4 as shown in Fig. 2). Despite some amount of statistical noise in the QMC calculation, it is clear that the RDG computed within DFT is in excellent quantitative agreement with QMC. Importantly, QMC and DFT both show, to the same quantitative extent, depletion of the RDG in the middle of the complex due to the interaction of the densities. The RDG peaks at a relatively large value of approximately 2~3 before diminishing to zero in the valley.

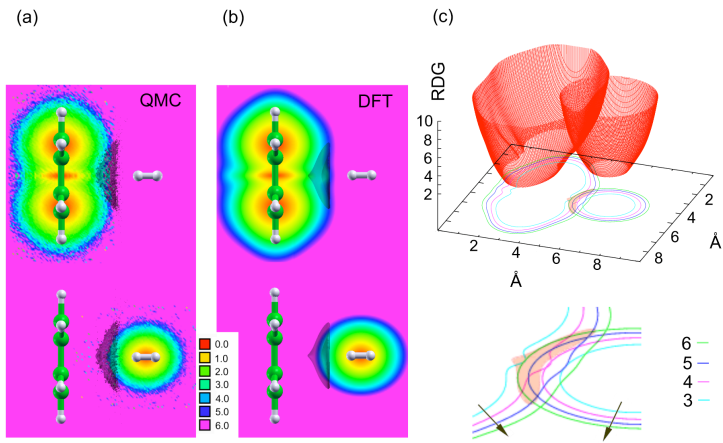

**Figure 3.** (a, b) Reduced Density Gradient (RDG) from R-QMC and DFT-PBE for the non-interacting densities at the fixed geometry. The shaded region in the middle indicates the "valley" region where the interacting densities show RDG less than 2 as in Figure 2. The RDG are plotted on the $C_{6v}$ symmetry plane containing two carbon atoms, showing the RDG valley region formed in the middle. (c) The contour plot is also shown for the DFT-PBE case.

We also compute the change in the RDG due to the interaction by computing the RDG of non-interacting densities in this valley region at the fixed benzene-$H_2$ complex geometry. A comparison of the QMC and DFT-PBE is shown in Figure. 3. Although the RDG appears to increase slightly faster for QMC than DFT-PBE, they show essentially the same quantitative features. For both cases, the RDG of the non-interacting benzene and $H_2$ were calculated to be greater than 3 in the valley region and as large as 6~7 where the RDG of the *interacting* density diminishes.

These observations reveal an important correlation of the binding energies observed in Fig 1 to the strength of exchange energy density enhanced within different exchange functionals for large RDG values. The presently discussed LDA and GGAs have exchange energy of the following form,

$$E_X[n] = \int d^3 r\, e_X^{Unif}(n(r)) F_X(s(r)),$$

where $e_X^{Unif}(n(r))$ is the exchange energy density of the uniform electron gas and $F_X(s(r))$ is the so-called exchange enhancement factor. This exchange enhancement factor determines the weight of the exchange energy density contribution as a function of the RDG, $s(r)$ at the given spatial point.

A major part of the exchange energy change due to the interaction results from the valley region, and it is to linear order given by changes in the exchange energy density and the enhancement factor. While the electron density is approximately the same for different XC approximations and thus also the exchange energy density ($\propto n^{4/3}$), the exchange enhancement factor behavior upon interaction deviates substantially among different XC approximations for the observed RDG change from large values ($3 \lesssim s \lesssim 6$) to small values ($0 \lesssim s \lesssim 2$). Among exchange functionals considered here, the exchange enhancement factor differs significantly as the RDG becomes larger despite the fact that they are essentially identical near $s = 0$ (See Supplementary Materials). Although the "exact" exchange enhancement factor cannot be constructed by remaining within GGA [21], there are a few mathematical constraints derived from physical considerations. The B88 exchange enhancement factor exceeds the upper bound imposed *locally* to satisfy the well-known Lieb-Oxford bound [22]. The PBE and PW91 functionals obey the bound and exhibit similar behavior until $s \sim 4$ where the PW91 curve begins to deviate significantly downward. All the GGA functionals here follow closely the asymptotic expansion suitable for exchange energies of free atoms [23]. They all, however, consequently violate the density gradient expansion condition [24]. The exchange enhancement factor in LDA is equal to unity by definition for all RDG values; therefore at a relatively large RDG value, the LDA exchange energy integrand is much smaller in magnitude than that in GGA exchange approximations.

This observation points to the importance of the enhancement factor behavior at large RDG among different exchange approximations for the significant difference observed in the binding energy for different XC approximations (Fig.1). The Slater exchange of LDA gives less weight for exchange energy density where the RDG is large compared to GGA exchanges, while the B88 exchange gives more weight than PBE exchange among them. This has a dominant effect regarding the extent to which different exchange approximations favor weak interactions as observed in our calculations.

Covalent interactions between molecules are generally characterized by a significant overlap of constituent densities where the RDG is small (close to the nuclei), thus most of the exchange energy *change* due to the interaction in covalently-bound systems results from the region where the RDG is quite small. It has been shown previously that the RDG in the range of $0 < s < 3$ is important for describing the atomic-shell structure [25], which is essential for describing the hybridization associated with covalent interactions (chemical bond formation). However, when the interaction is characterized *dominantly* by a small overlap of density tails, the exchange energy change results from the region where constituent densities have large RDG. This analysis also explains the previous observation by Zhang, et al. that for noble atom dimers the approximate exchange functional was found play an important role, noting the strong dependence on the large RDG values [26].

In order to assess the importance of varying exchange enhancement factor behaviors in DFT for describing weak interactions, we computed the binding energy of the benzene-$H_2$ complex using the LDA (Slater), B88, PBE, and PW91 enhancement factor for exchange approximation *combined with the PBE correlation approximation* (Table 1), given that the PBE XC approximation appears to behave the most accurately among them [27]. We found that this binding energy with Slater exchange is considerably large at 0.231 eV while with B88 exchange the complex is *not* bound. With PW91 exchange, this "PBE correlation" binding energy is 0.032 eV (quantitatively the same as the PW91 binding energy), which is quite similar to the PBE value of 0.026 eV. Hartree-Fock exchange energy with the PBE correlation energy would give a binding energy of 0.043 eV for comparison, closer to the QMC-PBE value of 0.058(6) eV.

These results taken together show *how* approximate exchange plays a critical role in the DFT description of weak interactions, impacting the binding energy significantly. The electron density of weakly-interacting systems in DFT calculations appear to be in good agreement with that from accurate reptation QMC calculation. At the same time, routine QMC calculations are computationally demanding and obtaining the energy derivatives for geometry optimization and molecular dynamics simulations remains challenging [28]. Therefore, it is of interest to improve the DFT calculation by understanding existing shortcomings of the widely-used XC approximations. Although the exact exchange energy within the DFT-KS scheme can be obtained through the optimized effective potential approach [29], improving the exchange enhancement factor by remaining within a GGA form is computationally attractive for a large class of systems.

The most straightforward improvement might be achieved by tailoring the large reduced density gradient behavior of the exchange enhancement factor for weak interactions (as done recently for solid-state/surface systems [30]), which is unlikely to affect noticeably the descriptions of other types of interactions such as covalent binding. In this Rapid Communication, our goal has been to improve our fundamental understanding of the shortcomings of the XC approximations for weak interactions, using accurate QMC calculations. The developments of XC approximations have followed largely two distinct approaches of either empirically fitting to available experimental data or of more formally satisfying known physics at certain limits. Continuing advancements of QMC methodologies in computing exact behaviors of the electron densities under various *realistic* environments might provide us another alternative approach for improving XC approximations from first principles.

**Acknowledgement**

We thank Lucas K. Wagner for fruitful discussions on quantum Monte Carlo methodologies. This work was performed under the auspices of the National Science Foundation by University of California, Berkeley under grant no. 0425914.

Table 1. Binding energies (BE) computed for Benzene-$H_2$ complex using PBE correlation and exchange energy with the enhancement factor from the approximations listed. The comparison to the binding energy using PBE correlation with Hartree-Fock (HF) exchange is also listed.

| Fx | Slater | B88 | PBE | PW91 | HF |
|---|---|---|---|---|---|
| BE (eV) | 0.231 | NA | 0.026 | 0.032 | 0.043 |